# THz Transient Photoconductivity of the III-V Dilute Nitride GaP$_y$As$_{1-y-x}$N$_x$.

J. N. Heyman[1], E. M. Weiss[1], J. R. Rollag[1], K. M. Yu[2,3], O. D. Dubon[2,4],

Y. J. Kuang[5], C. W. Tu[5], W. Walukiewicz[2]

[1.] Physics Department, Macalester College, St. Paul, MN 55105, USA. Email: heyman@macalester.edu.
[2.] Materials Sciences Division, Lawrence Berkeley National Laboratory, Berkeley, CA, 94720, USA.
[3.] Department of Physics, City University of Hong Kong, Hong Kong SAR, China.
[4.] Department of Materials Science and Engineering, University of California, Berkeley, California 94720, USA.
[5.] Department of Physics, University of California, San Diego CA 92093, USA.

THz Time-Resolved Photoconductivity is used to probe carrier dynamics in the dilute III-V nitride GaP$_{0.49}$As$_{0.47}$N$_{0.036}$. In these measurements a femtosecond optical pump-pulse excites electron-hole pairs, and a delayed THz pulse measures the change in conductivity. We find the photoconductivity is dominated by localized carriers. The decay of photoconductivity after excitation is consistent with bimolecular electron-hole recombination with recombination constant $r$ = 3.2±0.8·10$^{-8}$ cm$^3$/s. We discuss the implications for applications in solar energy.

## I.  INTRODUCTION

Dilute III-V nitride semiconductor alloys (III-V$_{1-x}$-N$_x$ with $x$<5%) are attractive for solar energy applications because their bandgaps are widely tunable with nitrogen concentration.[1] In$_y$Ga$_{1-y}$As$_{1-x}$N$_x$ has been used to create ~1eV bandgap sub-cells for multi-junction solar cells that are lattice-matched to GaAs.[2] While the internal quantum efficiency and output voltage of In$_y$Ga$_{1-y}$As$_{1-x}$N$_x$ solar cells were initially poor due to short carrier diffusion lengths in these materials, higher internal quantum efficiencies were achieved in the alloy system In$_y$Ga$_{1-y}$As$_{1-x-z}$N$_x$Sb$_z$.[3] Recently[4] three-junction solar cells that incorporate ~1eV bandgap In$_y$Ga$_{1-y}$As$_{1-x-z}$N$_x$Sb$_z$ sub-cells have been reported with concentrator efficiencies as high as 43.5%. A five-junction cell incorporating In$_y$Ga$_{1-y}$As$_{1-x}$N$_x$ has also been demonstrated.[5]

GaAs$_{1-x}$N$_x$, GaP$_y$As$_{1-x-y}$N$_x$ and other dilute III-V nitrides have also been proposed for use in intermediate band solar cells (IBSCs). Originally proposed by Luque and Marti,[6] IBSCs possess a band of delocalized electron states in the energy gap between the conduction and valance bands. In an IBSC solar photons excite electrons either directly from the valence band to the conduction band ($VB/E_+$), or sequentially with one photon promoting an electron from the valance band to the intermediate band ($VB/E_-$), and a second exciting it to the conduction band ($E_-/E_+$). In principle, IBSC can achieve efficiencies comparable to triple-junction solar cells in a single-junction device. In dilute III-V nitrides, interaction between the conduction band states and the nitrogen defect levels splits the conduction band into the lower energy ($E_-$) band and higher energy ($E_+$) band required for an IBSC. GaAs$_{1-x}$N$_x$,[7] GaSb$_y$As$_{1-x-y}$N$_x$,[8] and GaP$_y$As$_{1-x-y}$N$_x$[9-11] have been investigated for use in IBSC's, and the quarternary alloy GaP$_{0.4}$As$_{0.58}$N$_{0.02}$ is predicted to produce a nearly optimum match to the solar spectrum. However, bulk intermediate band solar cells produced to date exhibit low output voltages and low internal quantum efficiencies due to short carrier diffusion lengths in the $E_+$ band and low carrier lifetimes in the $E_-$ band. Estimates of the $E_-$ band carrier lifetime required for efficient IBSC's range from hundreds of nanoseconds to microseconds[12] because electrons promoted to the intermediate band must be excited into the $E_+$ band before they relax to the valence band.

Despite the flexibility of III-V dilute nitrides, high defect densities and short minority carrier diffusion lengths limit the efficiency of proposed devices. Carrier lifetimes in dilute III-V nitrides are strongly effected by nitrogen doping. Time Resolved Photoluminesence (TRPL) measurements at room temperature[13] in $n$-doped GaAsN found a carrier recombination rate proportional to the donor density. This behavior is typically associated with band to band electron-hole recombination, but the rate constant reported, $r = 4.6 \cdot 10^{-9} \text{cm}^3/\text{s}$, was an order of magnitude larger than that observed for radiative recombination in GaAs. Carrier lifetimes obtained by TRPL at low temperature are strongly dependent on luminescence energy.[14] Fast (0.35ns) non-radiative recombination was observed at shorter wavelengths corresponding to free-exciton recombination, while lifetimes an order of magnitude longer were observed at longer wavelengths associated with recombination of localized excitons. Time-resolved photoconductivity measurements[15] of $p$-GaAsN showed that a component of the photoconductivity decays on a microsecond decay time, and the authors suggested that nitrogen concentration fluctuations produced type II potential barriers leading to charge segregation and long recombination times.

Transient photoluminescence measurements of GaP$_y$As$_{1-x-y}$N$_x$ by Baranowski, *et. al.*[9]

reported $E_-/VB$ luminescence that decayed with characteristic times of ~0.1ns at room temperature, and ~100ns at $T$<100K. The authors' model included rapid electron-hole recombination at deep traps at room temperature, and long exciton lifetimes from excitons bound at shallow traps at low temperature.

We recently reported[16] room temperature carrier lifetime measurements in two $GaP_yAs_{1-x-y}N_x$ samples using transient absorption. An optical pump pulse excited carriers from the valance band to the $E_+$ and $E_-$ bands, leading to partial saturation of the $VB/E_-$ and $VB/E_+$ transitions as well as induced absorption of the $E_-/E_+$ transition. In $GaP_{0.32}As_{0.67}N_{0.01}$ we measured a 23ps recombination lifetime for the $E_+$ band. Carrier recombination from the intermediate band was non-exponential with a recombination rate proportional to the product of the electron and hole populations ($dn/dt = -r \cdot np$). We measured a recombination constant of $r = 2 \cdot 10^{-8}$ cm$^3$/s in $GaP_{0.32}As_{0.67}N_{0.01}$ and $r = 3.5 \cdot 10^{-8}$ cm$^3$/s in $GaP_{0.49}As_{0.47}N_{0.036}$, values approximately two orders of magnitude larger than predicted for radiative recombination in this material.

Nitrogen doping also strongly effects carrier transport in III-V semiconductors. Variable temperature Hall measurements in $In_yGa_{1-y}As_{1-x}N_x$ with $x$=2% by Kurtz, *et. al.*,[17] show room temperature electron mobilities of order 100 – 300 cm$^2$/Vs. They also find the electron and hole mobilities to be thermally activated, consistent with transport across potential barriers arising from fluctuations in nitrogen composition. Spectrally resolved PL and photoconductivity show a Stokes shift between absorption and luminescence consistent with carrier localization at low temperatures.

Time-resolved THz photoconductivity is particularly relevant to the characterization of solar cell materials because it measures photoconductivity directly. Time-resolved THz photoconductivity measurements of $GaAs_{1-x}N_x$ and $GaAs_{1-x-y}N_xBi_y$ by Cooke[18] show strong suppression of electron mobility with nitrogen doping. They found that the frequency-dependence of the photoconductivity of $GaAs_{1-x}N_x$ cannot be described by the Drude model. Instead the conductivity is suppressed at low frequencies due to charge localization.

In this work we report time-resolved THz photoconductivity measurements of $GaP_{0.49}As_{0.47}N_{0.036}$. We measure recombination rates that are consistent with our previous results from transient absorption. The frequency dependence of the conductivity indicates that the photo-carriers are weakly localized.

## II. EXPERIMENT

We report time-resolved photoconductivity measurements in $GaP_{0.49}As_{0.47}N_{0.036}$. Epitaxial $GaP_yAs_{1-x-y}N_x$ layers were grown on a GaP substrate in a modified Varian Gen-II molecular beam epitaxy system which has been described elsewhere.[10] Starting at the substrate, the structure consists (Figure 1a) of a 0.3-μm-thick GaP buffer layer, followed by a 1.5-μm-thick linearly graded $GaP_yAs_{1-y}$ layer in which $y$ increased from zero to the final value, and an 0.5-μm thick $GaP_{0.49}As_{0.51}$ layer. The 0.5-μm thick $GaP_{0.49}As_{0.47}N_{0.036}$ active layer was then grown using RF-plasma activated nitrogen. The sample structure was confirmed by Rutherford Back-Scattering (RBS), which determined the concentrations of Ga, As and P as a function of depth. These samples have previously been characterized by optical absorption and reflection, time-resolved luminescence, and transient absorption.[16]

The electronic band structure of the material was calculated using the band anticrossing model (BAC).[19,20] This simple model has been shown to accurately predict the bandgap and carrier effective mass in III-V dilute nitride semiconductors. The BAC predicts that the interaction between the extended conduction band states of the host material and the localized energy states of the isoelectronic nitrogen impurities splits the conduction band into distinct upper and lower bands $E_+$ and $E_-$ separated by a gap (Figure 1b). The BAC model also accurately predicts near band-edge optical absorption in our samples.[16]

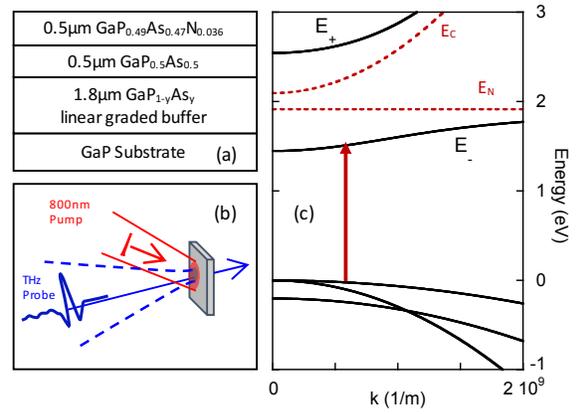

Figure 1. (a) Sample structure. (b) Optical pump, THz probe transient conductivity measurement. (c) Band structure of our $GaP_{0.49}As_{0.47}N_{0.036}$ sample. Solid lines: band anti-crossing model, dashed lines: uncoupled conduction band and the N impurity level positions, arrow: pump photon.

Samples were mounted epi-side down on quartz slides with optical epoxy and the substrates were thinned to ~50 μm. For THz measurements, a 1mm thick [100] GaP chip was optically contacted to the substrate to suppress THz standing waves. Time-resolved conductivity measurements were performed with an optical pump, THz probe system based on a FemtoLasers XL500 chirped pulse oscillator (800nm center wavelength, 5MHz repetition rate, 0.5μJ pulse energy, 50fs pulse-width). The laser pulses were divided into pump, THz generation and probe beams, and each beam incorporated an independent delay stage. The THz generation beam excited a biased photoconductive switch to generate THz pulses which were then collected and focused onto the sample. The transmitted THz beam was focused onto a 1mm-thick ZnTe crystal, and the THz electric field amplitude was measured by the probe beam using the electro-optic effect. The photoconductive emitter bias was modulated and the signal was recovered with a lock-in amplifier.

Scanning the THz beam delay at fixed probe delay allows measurement of the THz pulse waveform. The Fourier transform of the waveform yields a complex single beam spectrum $\tilde{E}(\omega)$. The transmission spectrum of the sample is the ratio of single beam spectra of the sample and a reference $\tilde{t}(\omega) = \tilde{E}_S(\omega)/\tilde{E}_R(\omega)$. The THz conductivity was determined from the transmission spectrum. We treated the epitaxial $GaP_yAs_{1-x-y}N_x$ layer as a 2D conducting sheet at the interface between the GaP substrate and the quartz. When internal reflections are ignored, the ratio of the transmission in the presence of the pump to the transmission with the pump blocked is:

$$\frac{\tilde{t}}{t_0} = \frac{n_2 + n_3}{n_2 + n_3 + \mu_0 c \tilde{\sigma}} \quad , \tag{1}$$

where $t_0$ is the amplitude transmission with $\tilde{\sigma} = 0$ (pump blocked), $n_2$ is the index of refraction of GaP and $n_3$ is the index of quartz.

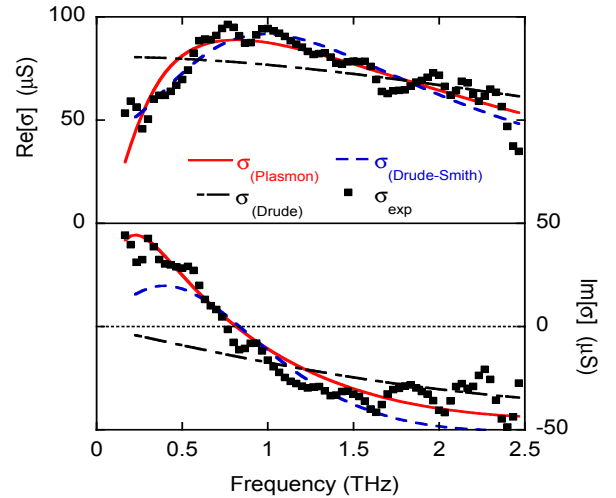

Figure 2. Real (top) and imaginary (bottom) photoconductivity versus frequency measured 7.5ps after photoexcitation. Solid line (red online) is the best fit to the Plasmon model. Dashed line (blue online) is the best fit Drude-Smith model. Dashed-dotted line (black) is the best fit Drude model.

In our time-resolved measurements, the pump pulse excited the samples at an incident photon flux $\phi \sim 10^{13} cm^{-2}$ at 800nm. The pump beam was chopped and the signal was recovered using lock-in amplifiers locked to the chopper frequency and the bias modulation frequency in series. We scanned the THz delay at a fixed pump delay relative to the probe to determine the change in the THz transmission spectrum due to the pump excitation. A series of measurements was used to map out the transmission spectrum as a function of delay. We found that the frequency-dependence of the conductivity spectrum was approximately independent of pump delay for delays

≥2.5 ps, so that a measurement of the THz signal at any point on the THz waveform yields a signal proportional to conductivity. This allowed us to set the THz delay to the peak signal point and sweep the pump delay to perform a fast and robust measurement of the decay of photoconductivity with time after photoexcitation.

## III. RESULTS

Our $GaP_{0.49}As_{0.47}N_{0.036}$ sample exhibits THz photoconductivity under 800nm excitation while a $GaP_{0.49}As_{0.51}$ nitrogen-free test sample and a bare GaP substrate do not. Figure 2 shows the conductivity versus frequency of our sample over the frequency range $f$ = 0.2 – 2.5THz, measured 7.5ps after photoexcitation. The real and imaginary parts of the complex conductivity were determined from the pump-induced transmission change using equation (1). The real part of the conductivity rises with increasing frequency to a maximum at ~0.9 THz and then decreases slowly towards higher frequencies. The imaginary part of the conductivity is positive at low frequencies and changes sign at ~0.9THz. The solid and dashed lines in this figure are model fits and are discussed below.

Figure 3 shows photoconductivity spectra obtained at a range of delays from -2.5ps to +1060ps. Photoconductivity vanishes at negative delays. The shape of the conductivity spectrum evolves rapidly during the first few picoseconds following pump excitation. For pump delays ≥4ps, the frequency-averaged magnitude of the conductivity decays with increasing delay and is almost indistinguishable from noise 1ns after pump excitation. At pump delays between 4ps and 350ps the shape of the real conductivity spectrum is approximately constant, rising with increasing frequency to a maximum near 0.9THz and then decreasing slowly with increasing frequency. The sign change in the imaginary component of the conductivity is clearly visible for delays between 4ps - 40ps, but becomes unobservable at larger delays. The magnitude of the imaginary component of the conductivity decays into noise for delays ≥ 350ps.

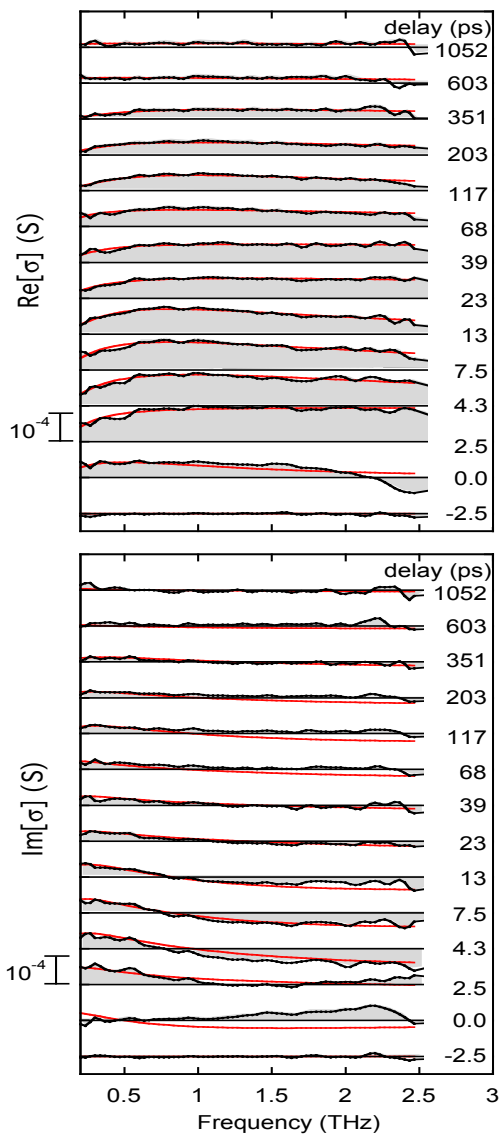

Figure 3. Real (top) and imaginary (bottom) photoconductivity spectra measured at delays ranging from -2.5ps to 1052ps after photoexcitation. Spectra are offset vertically for clarity. Black solid lines are experimental data, solid lines (red online) are fits to the Plasmon model.

## IV. DISCUSSION

The frequency dependence of the THz conductivity probes charge localization and scattering in our sample. The three simplest models used to describe charge transport in similar materials are[21] the Drude model, the Drude-Smith model and the Bound-Carrier or Plasmon model. The Drude model assumes that carriers scatter randomly with a single scattering time $\tau$ and yields a frequency-dependent conductivity:

$$\sigma_{Drude} = \frac{\sigma_0}{1+i\omega\tau}, \qquad (2)$$

where $\sigma_0 = ne^2\tau/m^*$ is the DC conductivity. The Drude model adequately describes the conductivity of many homogeneous conducting materials like metals and doped semicoductors. The Drude-Smith models the conductivity spectra of more complex materials. It assumes that a fraction of the carriers are coherently backscattered and yields a frequency-dependent conductivity:

$$\sigma_{Drude-Smith} = \frac{\sigma_0}{1+i\omega\tau}\left(1-\frac{c}{1+i\omega\tau}\right), \qquad (3)$$

where $c$ is a fitting parameter that can take on values between 0 and -1. This empirical model often fits experimental conductivity spectra well, although the value of the $c$ parameter obtained is difficult to associate with any physical property of the material. The Plasmon model assumes that the carriers are localized by a harmonic potential with a resonance frequency $\omega_0$. The conductivity versus frequency is then given by

$$\sigma_{Plasmon} = \frac{\sigma_0}{1+i\omega\tau(1-\omega_0^2/\omega^2)}, \qquad (4)$$

where $\sigma_0 = ne^2\tau/m^*$ is now the conductivity at $\omega = \omega_0$. This model can describe inhomogeneous materials in which carriers are localized into small conducting domains.

The Plasmon model gives the best fit to our measurements. Comparing our data to best-fit model spectra at delay of 7.5ps (Figure 2), one can see that the Drude model gives a poor fit, while the Drude-Smith and Plasmon models fit the data much more closely. In particular, the latter models reproduce the sign change observed in the imaginary component of the conductivity which is not captured by the Drude Model. We computed reduced chi-squared statistics for the entire data set by fitting spectra obtained at each delay value to each model. The Plasmon model gives the best fit to the data ($\tilde{\chi}^2 = 2.3$), followed by the Drude-Smith model ($\tilde{\chi}^2 = 3.7$) and the Drude model ($\tilde{\chi}^2 = 7.3$). The remainder of our analysis uses the Plasmon model to fit our data, and best-fit spectra obtained with the Plasmon model are plotted together with the experimental data in Figure 3.

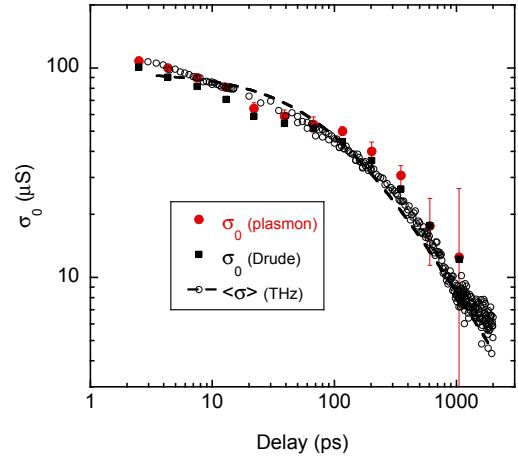

Figure 4. Conductivity coefficient $\sigma_0$ extracted from the Plasmon model (solid circles, red online) and the Drude model (black squares) versus delay. Open black circles show the conductivity estimated from single-point THz measurements. The dashed line shows a fit to the electron-hole recombination model discussed in the text.

The best-fit values of $\sigma_0$ as a function of delay obtained from the Plasmon model are plotted in Figure 4. Error bars were estimated using a Monte Carlo simulation with synthetic data sets.[22] The best-fit $\sigma_0$ values are robust: they agree with $\sigma_0$ values obtained by fits to the Drude model within ~20%, and they closely track approximate conductivity values found by measuring the pump-induced change in THz transmission at the peak signal point as a function of pump-delay (see Figure 4). The best-fit values of the parameters $\tau$ and $\omega_0$ are approximately constant at delays between 4ps - 13ps but are poorly constrained at larger delays. We have no clear evidence for systematic variation in $\omega_0$ and $\tau$ with delay and estimate uncertainty-weighted average values for these parameters $\omega_0/2\pi = 0.93 \pm 0.03$THz and $\tau = 37 \pm 5$fs.

Because $\omega_0$ and $\tau$ do not appear to vary significantly with delay, the conductivity values obtained by measuring the pump-induced change in THz transmission at the peak signal point track the photo-carrier density (Figure 4 and also a preliminary report by Weiss, *et al.* [23]). We fit our results with an electron-hole bimolecular recombination model $dn/dt = -r \cdot np$ which, for our undoped samples ($n$ = $p$) predicts the carrier density versus time:

$$n(t) = \frac{n_0}{1 + r n_0 t} \quad . \tag{5}$$

Here $n_0 = 3.1 \pm 0.8 \cdot 10^{17} \text{cm}^{-3}$ is the free electron concentration immediately after photoexcitation estimated from the absorbed pump photon flux. The model provides an excellent fit to our data and we obtain a rate constant $r$ = 3.2±0.8·10$^{-8}$ cm$^3$/s, where the uncertainty in $r$ is dominated by the uncertainty in the photon flux. These results are consistent with our previous transient absorption measurements on the same sample[16] carried out at photoelectron concentrations of $10^{18} \text{cm}^{-3}$ to $10^{19} \text{cm}^{-3}$, so that the bimolecular recombination model with a *single free parameter* is able to describe independent experiments carried out at widely different photo-carrier densities. In contrast, although a stretched exponential model (two free parameters) and a double exponential decay model (three free parameters) are each able to fit our THz data well, neither can also fit the transient absorption data with a consistent set of parameters.

Our conductivity spectra indicate photo-carrier localization which may result from inhomogeneity in nitrogen concentration or carrier trapping at defects. While the low value of the resonance frequency ($\hbar\omega_0 \sim$4meV) indicates that the localization is weak, these results indicate that the DC mobility of the photocarriers should be much smaller than the effective mobility at THz frequencies, $\mu_{eff} = \sigma_0/ne \sim 30 \frac{\text{cm}^2}{\text{Vs}}$. Low carrier mobility and carrier localization associated with nitrogen doping have been widely reported and represent a barrier to more widespread applications.

## V.    CONCLUSIONS

We have used Time-Resolved THz Spectroscopy to investigate carrier dynamics in the III-V dilute nitride GaP$_{0.49}$As$_{0.47}$N$_{0.036}$. We find that bimolecular electron-hole recombination describes carrier recombination in this material with a recombination constant of $r \sim$ 3.2±0.8·10$^{-8}$ cm$^3$/s. This value is approximately 100x larger than the predicted radiative recombination time for this material,[16] suggesting that the recombination is mediated by defects. This recombination mechanism strongly constrains applications of this material for intermediate band solar cells. IBSC designs typically use n-doping (~10$^{19}$cm$^{-3}$) to place the Fermi level in the middle of the intermediate band to boost the intermediate to conduction band absorption coefficient. Bimolecular recombination with this *r* value in such heavily doped material would produce single-exponential decay with ~10ps minority carrier lifetimes ($\tau \sim 1/rn$). This recombination mechanism is not problematic for carrier generation in the depletion layer of a *pn*-junction solar cell, although the suppressed DC mobility of photo-carriers that we report would negatively impact the carrier diffusion length. These material problems must be addressed before this GaP$_y$As$_{1-y-x}$N$_x$ will be widely useful for solar energy applications.

## VI.     ACKNOWLEDGEMENTS

Work performed in the Electronic Materials Program was supported by the Office of Science, Office of Basic Energy Sciences, of the U.S. Department of Energy under Contract No. DE-AC02-05CH11231. KMY acknowledges the support of the General Research Fund of the Research Grants Council of Hong Kong SAR, China, under project number CityU 11303715. Samples were synthesized in Prof. Tu's laboratory at UC San Diego by Y.J.K and C.W.T. The experiments were carried out by J. H., J. R and E. W. at Macalester College. The interpretation of the experimental results was done by J. H. and E. W. The other authors characterized the samples and provided guidance on this project.

**Figure Captions**

Figure 1. (a) Sample structure. (b) Optical pump, THz probe transient conductivity measurement. (c) Band structure of our GaP$_{0.49}$As$_{0.47}$N$_{0.036}$ sample. Solid lines: band anti-crossing model, dashed lines: uncoupled conduction band and the N impurity level positions, arrow: pump photon.

Figure 2. Real (top) and imaginary (bottom) photoconductivity versus frequency measured 7.5ps after photoexcitation. Solid line (red online) is the best fit to the Plasmon model. Dashed line (blue online) is the best fit Drude-Smith model. Dashed-dotted line (black) is the best fit Drude model.

Figure 3. Real (top) and imaginary (bottom) photoconductivity spectra measured at delays ranging from -2.5ps to 1052ps after photoexcitation. Spectra are offset vertically for clarity. Black solid lines are experimental data, solid lines (red online) are fits to the Plasmon model.

Figure 4. Conductivity coefficient $\sigma_0$ extracted from the Plasmon model (solid circles, red online) and the Drude model (black squares) versus delay. Open black circles show the conductivity estimated from single-point THz measurements. The dashed line shows a fit to the electron-hole recombination model discussed in the text.